\documentclass[useAMS,usenatbib]{mn2e}

\def\h2 {H$_2$\,}
\def\Jcrit {J_\mathrm{crit}}
\def\ccm {\,\mathrm{cm}^{-3}}
\def\TreeCol {{\sc treecol}\,}

\usepackage{amsmath}
\usepackage{amssymb}
\usepackage{graphicx}

\title[Improving H$_2$ self-shielding]{How an improved implementation of H$_2$ self-shielding influences the formation of massive stars and black holes}

\author[T.\ Hartwig et al.]
{\parbox{\textwidth}{
Tilman Hartwig$^{1,2}$\thanks{E-mail: hartwig@iap.fr}, Simon C.\ O.\ Glover$^{3}$, Ralf S.\ Klessen$^{3,4,5}$, Muhammad\ A.\\
Latif$^{1,2}$ and Marta Volonteri$^{1,2}$\\
}\\
$^{1}$Sorbonne Universit\'es, UPMC Univ Paris 06, UMR 7095, Institut d'Astrophysique de Paris, F-75014, Paris, France\\
$^{2}$CNRS, UMR 7095, Institut d'Astrophysique de Paris, F-75014, Paris, France\\
$^{3}$Universit\"at Heidelberg, Zentrum f\"ur Astronomie, Institut f\"ur Theoretische Astrophysik, Albert-Ueberle-Str.\ 2,\\
D-69120 Heidelberg, Germany\\
$^{4}$Department of Astronomy and Astrophysics, University of California, 1156 High
Street, Santa Cruz, CA 95064, USA\\
$^{5}$Kavli Institute for Particle Astrophysics and Cosmology, Stanford University,
SLAC National Accelerator Laboratory, Menlo Park,\\
CA 94025, USA}

\begin{document}


\pagerange{\pageref{firstpage}--\pageref{lastpage}} \pubyear{2015}

\maketitle

\label{firstpage}

\begin{abstract}
High-redshift quasars at $z>6$ have masses up to $\sim\,10^9\,\mathrm{M}_\odot$. One of the pathways to their formation includes direct collapse of gas, forming a supermassive star, precursor of the black hole seed. The conditions for direct collapse are more easily achievable in metal-free haloes, where atomic hydrogen cooling operates and molecular hydrogen (H$_2$) formation is inhibited by a strong external UV flux. Above a certain value of UV flux ($\Jcrit$), the gas in a halo collapses isothermally at $\sim10^4$\,K and provides the conditions for supermassive star formation. However, \h2 can self-shield, reducing the effect of photodissociation. So far, most numerical studies used the local Jeans length to calculate the column densities for self-shielding. We implement an improved method for the determination of column densities in 3D simulations and analyse its effect on the value of $\Jcrit$. This new method captures the gas geometry and velocity field and enables us to properly determine the direction-dependent self-shielding factor of \h2 against photodissociating radiation. We find a value of $\Jcrit$ that is a factor of two smaller than with the Jeans approach ($\sim\,2000\,J_{21}$ vs. $\sim\,4000\,J_{21}$). The main reason for this difference is the strong directional dependence of the \h2 column density. With this lower value of $\Jcrit$, the number of haloes exposed to a flux $>\Jcrit$ is larger by more than an order of magnitude compared to previous studies. This may translate into a similar enhancement in the predicted number density of black hole seeds.
\end{abstract}


\section{Introduction}
Observations of quasars at high redshifts indicate that supermassive black holes (SMBHs) of several billion solar masses were already assembled in the first billion years after the big bang \citep{fan03,fan06,willott10,venemans13,derosa14,wu15}. The current record holders are a bright quasar, which hosts a SMBH with a mass of $2 \times 10^9 \mathrm{M}_\odot$ at $z=7.085$ \citep{mortlock11}, corresponding to $\sim 800$ million years after the big bang, and a SMBH with $1.2 \times 10^{10} \mathrm{M}_\odot$ at $z=6.30$ \citep{wu15}. It is still unclear how these objects were able to acquire so much mass in this short period of time, which in turn raises questions about the formation mechanism and the involved physical processes. One possible explanation is the collapse of dense stellar clusters \citep{portegies04,omukai08,devecchi09,lupi14}.
Another scenario involves stellar mass seed black holes with masses up to a few hundred M$_\odot$ that are the remnants of Population III (Pop III) stars and then grow by mass accretion or mergers \citep{madau01,haiman01,volonteri03,yoo04,haiman06,pelupessy07,tanaka09,whalen12,madau14}. However, already a simple order of magnitude argument shows that this process involves some difficulties. Assuming accretion at the Eddington limit, the e-folding time is 50 million years \citep{milos09b}. In 800 million years, a seed black hole accreting at the Eddington limit can therefore grow by a factor of $e^{800/50} \simeq 9 \times 10^{6}$, and so to reach a mass of $2 \times 10^{9} \: {\rm M_{\odot}}$ by $z = 7.085$, it is necessary to start with a seed mass of $\sim 200 \: {\rm M_{\odot}}$. This is significantly larger than the mass of a typical Pop III stellar remnant \citep{clark11a,clark11b,greif11,stacy12,latif13b,hirano14,hartwig15}, but not yet completely ruled out.
Moreover, it is still an open question, how these high gas accretion rates could be sustained during the growth of the SMBH \citep{alvarez09,milos09a,johnson13,jeon14}.

A more promising formation scenario is the direct collapse of a protogalactic gas cloud, which yields black hole seed masses of $10^4-10^6 \mathrm{M}_\odot$ \citep{rees84,loeb94,bromm03,begelman06,volonteri10,shang10,schleicher10,choi13,latif13a,regan14b,latif14a,sugimura14,visbal14,agarwal14,latif15a,becerra15}. 
To form such a massive seed, a high mass inflow rate of $\gtrsim 0.1\,\mathrm{M}_\odot\,\mathrm{yr}^{-1}$ is required \citep{begelman10,hosokawa12,hosokawa13,schleicher13,ferrara14}. Sufficient conditions for such high mass inflow rates are provided in haloes with $T_\mathrm{vir} > 10^4$\,K in which gas fragmentation and star formation are suppressed during the collapse \citep{latif13a}. To avoid fragmentation, the gas has to be metal free and a strong radiation background has to photodissociate molecular hydrogen, which otherwise acts as a strong coolant. Under these specific conditions, the gas can only cool by atomic hydrogen and collapses monolithically to form a supermassive star (SMS), which later on forms a SMBH seed \citep{begelman10,hosokawa12,hosokawa13,inayoshi14a,inayoshi14b} or a quasi-star, which forms a stellar mass black hole that grows by swallowing its envelope \citep{begelman06,begelman08,ball11,schleicher13}. We will refer to this specific type of direct collapse as `direct collapse scenario' hereafter.

Based on the strength of the photodissociating radiation, the cloud either monolithically collapses close to isothermality, or is able to efficiently cool and to fragment. The main quantity that discriminates between these two different collapse regimes is the flux in the Lyman--Werner (LW) bands ($11.2--13.6$\,eV). This LW radiation is emitted by the first generation of stars and it is convenient to express the flux in units of $J_{21}=10^{-21}\,\mathrm{erg}\,\mathrm{s}^{-1}\,\mathrm{cm}^{-2}\,\mathrm{Hz}^{-1}\,\mathrm{sr}^{-1}$ (in the following, we use this convention without explicitly writing $J_{21}$). The so-called critical value $J_\mathrm{crit}$ sets the threshold above which a halo with $T_\mathrm{vir} > 10^4$\,K can directly collapse to a SMBH seed. Below this value, the gas is susceptible to fragmentation due to efficient H$_2$ cooling and the mass infall rates towards the centre are generally lower. However, the values for $\Jcrit$ quoted in the literature span several orders of magnitude from $\Jcrit=0.5$ \citep{agarwal15b} to as high as $\Jcrit \simeq 10^5$ \citep{omukai01,latif15a}. There are several reasons for this large scatter. First of all, the value of $J_{\rm crit}$ is highly sensitive to the spectral shape of the incident radiation field, with softer radiation fields leading to significant smaller values of $J_{\rm crit}$ \citep{shang10,sugimura14,agarwal15a,latif15a}. Secondly, one-zone calculations \cite[e.g.][]{omukai01} tend to yield lower values of $\Jcrit$ than determinations made using 3D numerical simulations. This is a consequence of the fact that $J_{\rm crit}$ depends to some extent on the details of the dynamical evolution of the gas, which are only approximately captured by one-zone calculations. This dependence on the gas dynamics also leads to $J_{\rm crit}$ varying by a factor of a few from halo to halo \citep{shang10,latif14a}.
Although there seems not to be one universal value of $J_\mathrm{crit}$ \citep{agarwal15b}, it is convenient to use this artificial threshold as a quantification of the direct collapse scenario to test the relevance of different physical processes. Once a process significantly affects the value of $\Jcrit$, it is very likely that it plays an important role in the formation of SMSs and SMBH seeds.

One of these important processes is H$_2$ self-shielding against LW radiation, which is generally expressed as a suppression factor $f_\mathrm{sh}$ to the H$_2$ photodissociation rate (see section \ref{sec:photodis}). There are several analytic expressions to calculate $f_\mathrm{sh}$ as a function of the H$_2$ column density and the gas temperature \citep{draine96,wolcott11,richings14}.
Neglecting self-shielding leads to a large change in $J_{\rm crit}$ \citep{shang10} and even among these analytic functions, the value of $\Jcrit$ varies by an order of magnitude \citep{latif14a,sugimura14}, which hence shows the importance of a correct treatment of this effect.
Another challenge is the proper determination of the effective \h2 column density for self-shielding, since it is either computationally very expensive or not very accurate. In this study, we want to test the effect of a more accurate \h2 self-shielding implementation on the direct collapse scenario. In contrast to previous studies, we determine the column densities self-consistently during the simulation and properly account for the Doppler shifts of spectral lines by velocity gradients, which reduce the effective column density. To do so, we use the {\sc treecol} algorithm developed by \citet{clark12} and extended by \citet{hartwig15} to calculate a spherical map of column densities around each cell.
This method is based on the hierarchical tree structure used to calculate the gravitational forces between fluid elements in the computational domain and therefore comes with only little additional cost. A more detailed description of this method is provided in section \ref{sec:photodis}.

The paper is organized as follows. In section \ref{sec:methods}, we describe the methods, including initial conditions, chemistry network, and the new implementation to determine self-shielding. In section \ref{sec:results}, we present the results together with an analysis of the differences between the self-shielding methods and the mass infall rates. We discuss the caveats in section \ref{sec:caveats} and conclude with a summary in section \ref{sec:summary}.

\section{Methodology}
\label{sec:methods}
In this section, we present our computational methods. First, we explain our initial conditions and refinement strategy. Then, we present the chemical network and our new approach for the determination of effective \h2 column densities for self-shielding.

\subsection{Initial conditions}
\label{sec:init}
We are interested in the collapse of the subset of metal-free haloes that is able to cool by atomic hydrogen. The cooling rate of H rises steeply around $T \simeq 10^4$\,K and expressed by the virial mass, we are interested in haloes with $M_\mathrm{vir} \simeq 10^7 \mathrm{M}_\odot$. In this study, we focus on the effect of different H$_2$ self-shielding implementations.
Consequently, we first run a cosmological dark matter only simulation and select the first haloes with a mass of $\sim 10^7\,\mathrm{M}_\odot$. Under the assumption that the value of $\Jcrit$ is mainly affected by the gas dynamics within the virial radius, this is a representative candidate for a metal-free halo that directly collapses to a SMBH seed.
We assume a flat $\Lambda$CDM Universe and use cosmological parameters presented by the \citet{planck14} with additional constraints from \textit{WMAP} polarisation at low multipoles, high-resolution cosmic microwave background data sets, and baryonic acoustic oscillations: $H_0=67.8\,\mathrm{km}\,\mathrm{s}^{-1}\,\mathrm{Mpc}^{-1}=h\,100\,\mathrm{km}\,\mathrm{s}^{-1}\,\mathrm{Mpc}^{-1}$, $\Omega _\Lambda = 0.69$, $\Omega _m = 0.31$, $\Omega _b = 0.048$, $n_s = 0.96$, $\sigma _8 = 0.83$, $Y_\mathrm{He}=0.25$. We create the initial density field at redshift $z=99$ in a periodic box of 1 Mpc\,$h^{-1}$ comoving with {\sc music} \citep{music}, which generates the displacements and velocities following second-order Lagrangian perturbation theory.

The simulations are performed with the hydrodynamic moving-mesh code {\sc arepo} \citep{arepo}, which combines the advantages of smoothed particle hydrodynamics (SPH) techniques and adaptive mesh refinement (AMR) codes. We first run a cosmological simulation with $\sim 2 \times 10^7$ Voronoi cells, which corresponds to a particle mass of $m_\mathrm{DM} = 5.1 \times 10^4 \mathrm{M}_\odot$.
We trace the target halo and a region of twice its virial radius to the initial conditions. In a second run, we refine this region of interest and also include gas, which leads to masses in the highest refined region of $m_\mathrm{DM}=100\,\mathrm{M}_\odot$ and $m_\mathrm{gas}=18\,\mathrm{M}_\odot$. This resolution is set to properly resolve the collapse of the gas up to densities of $n \simeq 10^{6}\,\mathrm{cm}^{-3}$. This value is chosen to cover the local thermodynamical equilibrium (LTE) of H$_2$ at around $n \simeq 10^{4}\,\mathrm{cm}^{-3}$, because above this value it is much easier to collisionally dissociate H$_2$ than at lower densities. Hence, once the gas reaches this value without building up a significant fraction of H$_2$, it is not going to manage to do so at higher densities either. \citet{regan15} study the effect of the dark matter mass resolution that is needed to properly resolve the collapse of haloes at high redshift. They find that for typical collapse scenarios with a moderate LW background, $m_\mathrm{DM}=100 \mathrm{M}_\odot$ is a sufficient resolution, whereas this minimum mass resolution even decreases for the higher LW backgrounds ($J_\mathrm{LW}=500$) that we want to study here. Consequently, our dark matter mass resolution is sufficient to properly resolve the collapse.

The finite box size of our simulations might distort the nonlinear effective coupling on the boxscale by not covering all relevant Fourier modes of the power spectrum \citep[see e.g.,][]{seto99}. However, the choice of our box size is well motivated for a cosmological representative selection of a $10^7 \mathrm{M}_\odot$ halo, because we do not want to draw high-precision cosmological probes from these simulations, but rather analyse the collapse behaviour of one specific halo. Another effect of the limited box size is the distortion of the large-scale tidal fields, which might affect the angular momentum of the haloes, according to tidal torque theory \citep[see e.g.,][]{fall80}. However, the angular momentum budget of haloes is dominated by local effects like mergers or the accretion of cold gas streams and only a minor contribution comes from cosmic tidal fields \citep{danovich12,dubois14,laigle15}. In any case, the effect of different implementation of H$_2$ self-shielding on the collapse dynamics should not be affected by the limited box size.

\subsection{Chemistry}
\label{sec:chem}
In the following section, we describe our chemical network and highlight the most important reactions and rate coefficients. A more extensive discussion of the relevant chemical processes for modelling direct collapse with a strong LW background can be found in \citet{glover15a,glover15b}. We apply a primordial chemistry network that is originally based on the work by \citet{glover07}, \citet{glover08}, and \citet{clark11a}. Since the deuterium chemistry does not affect the direct collapse scenario \citep{glover15a}, we only follow explicitly the evolution of H, He, H$_2$, H$^+$, H$^-$, H$^+_2$, He$^+$, He$^{++}$, and e$^-$.
\citet{glover15a} identified a minimal subset of reactions that must be included in the chemical model in order to determine $J_{\rm crit}$ accurately. We have made sure to include all of these reactions in our chemical network. Full details regarding our choice of reaction rate coefficients can be found in \citet{clark11a} and \citet{glover15a,glover15b}.

The collapse dynamics depends strongly on the abundance of molecular hydrogen, which is the dominant coolant for temperatures below $\sim 10^4$\,K. Molecular hydrogen is mainly formed via the two-step process:
\begin{align}
\label{eq:h2form}
\mathrm{H} + \mathrm{e}^-  &\rightarrow \mathrm{H}^- + \gamma\\
\mathrm{H} + \mathrm{H}^-  &\rightarrow \mathrm{H}_2 + \mathrm{e}^-,
\end{align}
and is primarily destroyed either by collisions with hydrogen atoms
\begin{equation}
\label{eq:h2col}
\mathrm{H}_2 + \mathrm{H} \rightarrow \mathrm{H} + \mathrm{H} + \mathrm{H},
\end{equation}
or by the so-called Solomon process \citep{field66,stecher67}:
\begin{equation}
\label{eq:h2dis}
\mathrm{H}_2 + \gamma \rightarrow \mathrm{H} + \mathrm{H},
\end{equation}
where a LW photon photodissociates the molecule by exciting it from the electric ground state into an excited electronic state. In $\sim 15 \%$ of the cases, the electrons do not decay into a bound state, but into the vibrational continuum of the ground state and thereby dissociate the molecule. For the Solomon process, we assume a spectrum of a blackbody with an equivalent temperature of $T_\mathrm{rad}=10^5$\,K (T5) that is cut off above $13.6$\,eV, because photons with higher energies are absorbed by the intergalactic medium. This choice represents the case in which the spectra are dominated by Pop~III stars. The second generation of stars is believed to be less massive and can be approximated by a blackbody spectrum with $10^4\,\mathrm{K}<T_\mathrm{rad}<10^5$\,K \citep{sugimura14,agarwal15a}. It is important to note that the value of $\Jcrit$ depends on the choice of the spectrum, because whereas a T5 spectrum mainly photodissociates the \h2 directly (equation \ref{eq:h2dis}), a spectrum with a cooler effective temperature dominantly prohibits \h2 formation by photodetachment of H$^-$ via
\begin{equation}
\label{eq:hmdis}
\mathrm{H}^- + \gamma \rightarrow \mathrm{H} + \mathrm{e}^-
\end{equation}
\citep{latif14a,sugimura14,agarwal15a,agarwal15b}. Since we want to focus on the effects of \h2 self-shielding, we will only consider the T5 spectrum, where this effect plays a dominant role. We also include dissociative tunnelling, which is discussed in \citet{martin96}. This process significantly contributes to the total collisional dissociation rate of \h2 and is therefore necessary for a proper treatment of primordial gas physics \citep{latif14a,glover15a}.

\subsection{H$_2$ self-shielding}
\label{sec:photodis}
Since one photon of the external radiation field can only photodissociate one H$_2$ molecule, a large column density of molecular hydrogen can protect the inner regions against photodissociation. This process is known as self-shielding. For the photodissociation of H$_2$, we use the rate coefficient \citep{glover07}
\begin{equation}
k=1.38 \times 10^{-12}  \frac{J_\mathrm{LW}}{J_{21}} f_\mathrm{sh}\,\mathrm{s}^{-1},
\label{eq:dissrate}
\end{equation}
where the factor $f_\mathrm{sh} \leq 1$ accounts for the effect of H$_2$ self-shielding with $f_\mathrm{sh} = 1$ in the optically thin limit. This rate coefficient corresponds to a normalisation of the radiation field of $J_\mathrm{LW} = 1$ at $12.87$\,eV, the middle of the LW bands. The exact treatment of H$_2$ self-shielding requires a full radiative scheme with line transfer of all the important lines in the LW bands and is therefore prohibitively expensive. However, the shielding factor can be approximately expressed as a function of the H$_2$ column density and the gas temperature. The latter enters because of the thermal broadening of spectral lines and due to the temperature-dependent excitation of different rotational levels of the \h2 molecule. \citet{draine96} study the structure of stationary photodissociation fronts and propose a self shielding factor of the form
\begin{equation}
f_\mathrm{sh} = \frac{0.965}{(1+x/b_5)^\alpha} + \frac{0.035}{(1+x)^{0.5}} \exp \left[ - \frac{(1+x)^{0.5}}{1180} \right],
\label{eq:db96}
\end{equation}
where $x=N_{\mathrm{H}_2} / 5 \times 10^{14} \mathrm{cm}^{-2}$ is the H$_2$ column density, $b_5 = b/10^5 \mathrm{cm}\,\mathrm{s}^{-1}$ is the scaled Doppler parameter of the molecular hydrogen, and $\alpha = 2$. This functional form accounts for the effect of line overlap and has been applied in many studies \citep{glover07,whalen08,gnedin09,shang10,glover10,christensen12,krumholz12}.

\citet{wolcott11} model the photodissociation and self-shielding of H$_2$ in protogalaxies with three-dimensional simulations, based on post-processing their output data.
They found that the formula by \citet{draine96} is only valid for cold or low density gas, in which only the lowest rotational states of H$_2$ are populated and propose the modification $\alpha = 1.1$ to equation (\ref{eq:db96}), which provides a better fit in dense gas for all relevant temperatures. A comparison of both functions \citep{latif14a,sugimura14} shows that the application of the newer formula by \citet{wolcott11} yields values of $\Jcrit$ that are up to an order of magnitude lower than those derived with the function by \citet{draine96}. Another functional form was proposed by \citet{richings14} who model shielding against UV radiation in the diffuse interstellar medium. They also include the effect of turbulent gas velocities, but they derive the self-shielding factor for a one-dimensional plane-parallel slab of gas. Although the fitting function by \citet{wolcott11} was derived in a three-dimensional simulation, it is also based on a static slab of gas.

In a more realistic scenario, however, we are interested in a collapsing cloud, where the relative velocities between infalling gas particles Doppler-shift the spectral lines. Due to this effect, an H$_2$ molecule can only shield other H$_2$ molecules whose relative velocity is smaller than its thermal velocity. Otherwise, the spectral lines are shifted too far and H$_2$ molecules do not contribute to the effective column density. To account for this effect, \citet{hartwig15} have implemented a new method for the determination of effective column densities in three-dimensional simulations. This method is based on the {\sc treecol} algorithm by \citet{clark12}, which directly sums up the individual mass contributions for the column density of each fluid element. With this information, we create spherical maps of the column density around each Voronoi cell, with 48 equal-area pixels based on the {\sc healpix} algorithm \citep{gorski05}. The number of 48 pixels is motivated by {\sc healpix}, which divides the sphere into 12 equal-area pixels, which can be subdivided into $2^N$ pixels each. Based on the work by \citet{clark12}, we chose $N=2$, since this value provides a sufficient angular resolution for most astrophysical applications.

Two characteristics of the code make \TreeCol highly useful for our purpose. First, it uses the tree structure, which is already present in the code to determine the gravitational force. Hence, the determination of column densities comes with only little additional computational cost. Secondly, we can directly compare the relative velocities of the particles $v_r$ that possibly contribute to the column density to the thermal gas velocity $v_\mathrm{th}$ of the particle for which we want to calculate the column density. Following \citet{hartwig15}, a particle only contributes to the effective column density, if
\begin{equation}
 v_r < 1.694 v_\mathrm{th},
 \label{eq:hart15}
\end{equation}
where $v_\mathrm{th}$ is the thermal gas velocity and the numerical factor is an extension of the so-called Sobolev approximation \citep{sobolev60} and takes the true line profile into account. Based on this criterion, we determine the effective column densities and calculate the shielding factors separately for all pixels with equation (\ref{eq:db96}) and the exponent $\alpha = 1.1$ by \citet{wolcott11}. The final shielding factor is the mean of 48 directional-dependent factors.

With this new approach, we can use formula (\ref{eq:db96}) with the \citet{wolcott11} exponent $\alpha = 1.1$, which was derived for static gas and extend it to collapsing gas clouds by defining the effective H$_2$ column densities based on the relative gas velocities. Since our approach automatically accounts for turbulent motions on scales above the spatial resolution, the Doppler parameter $b_5$ in formula (\ref{eq:db96}) does only include the thermal broadening. This approach cannot only be used for \h2 self-shielding, but also for many other radiative transfer processes that rely on the determination of column densities. The only requirement is that the resolution elements are stored in a tree-like structure, which is already the case in most codes that include self-gravity. This method for the determination of effective column densities is tested and explained in detail in \citet{hartwig15}.

Our method of computing effective \h2 column densities is valid as long as the main contribution comes from the core of individual lines that shield themselves. At high column densities, however, the Lorentzian contribution to the line profile becomes important and the corresponding damping wings should be taken into account for the determination of self-shielding. While this effect is negligible at small column densities, \citet{gnedin14} show that it should be taken into account for \h2 column densities of $N_{\mathrm{H}_2} > 10^{21}\,\mathrm{cm}^{-2}$. At these high column densities, the Doppler-shifts induced by relative velocities become less important and eventually the total column density contributes to self-shielding. A more detailed analysis of this effect is given in section \ref{sec:wings}, where we show that a correct treatment of the overlap of these damping wings changes the value of the self-shielding factor by less than $5\%$ and has no influence on the determination of $\Jcrit$. In addition, the use of effective column densities is computationally more efficient, since the velocity criterion imposed by equation (\ref{eq:hart15}) limits the amount of fluid elements that have to be projected. Therefore, we employ this method in our numerical simulations.

We also note that \citet{safranek12} use a non-local approach for the determination of \h2 column densities. They study the influence of LW radiation on star formation in the first galaxies and approximate the column density in the following way. For each computational cell, they calculate the column densities in six directions parallel to the coordinate axis. The smallest of these column densities is then used to calculate the self-shielding factor. They show that this is already an improvement, compared to local estimations of the column density. Similar techniques have also been used to study the effects of H$_{2}$ self-shielding in giant molecular clouds \citep[see e.g.][]{nl97,gm07a,gm07b}.

Most other previous simulations use an approximation for the H$_2$ column density, based on a characteristic length scale $L_\mathrm{char}$, and the assumption that the H$_2$ density is constant within this length and negligible beyond it. The column density is then given by $N_{\mathrm{H}_2}=n_{\mathrm{H}_2}L_\mathrm{char}$, where $n_{\mathrm{H}_2}$ is the local number density of molecular hydrogen. Assuming that the effect of self-shielding occurs only locally, many simulations \citep[e.g.][]{shang10,vanborm14,sugimura14,latif15a,latif15b,glover15a,glover15b,agarwal15b} use $L_\mathrm{char}=L_\mathrm{J}$ with the Jeans length
\begin{equation}
 L_\mathrm{J} = \sqrt{\frac{15 k_B T}{4 \pi G \mu m_p \rho}},
 \label{eq:Jeans}
\end{equation}
where $T$ is the temperature, $\mu$ the mean molecular weight and $\rho$ the gas density. Since this is a widely used approximation, we will compare our results obtained with \TreeCol to this formula.

\section{Results}
\label{sec:results}
In order to increase the statistical significance of our results, we create four independent sets of cosmological initial conditions. The side length of each of these boxes is 1Mpc\,$h^{-1}$, and we select one halo per box. As described in section \ref{sec:init}, the region of interest is refined for each box and we resimulate this set of cosmological zoom-in simulations.
At redshift $z=30$, we switch on the photodissociating background. In reality, the LW radiation increases with cosmic time and the time of its onset also influences the collapse of primordial gas clouds \citep{visbal14}. We use this simplification of an instantaneous onset to be able to focus on the different implementations of the H$_2$ self-shielding and make it comparable to most previous works in this field. As long as the H$_2$ abundance has enough time to reach an equilibrium, before the halo of interest starts with the run-away collapse, the results are unaffected by the choice of this redshift. This criterion is fulfilled in all our simulations.

\begin{figure}
\centering
\includegraphics[width=0.47\textwidth]{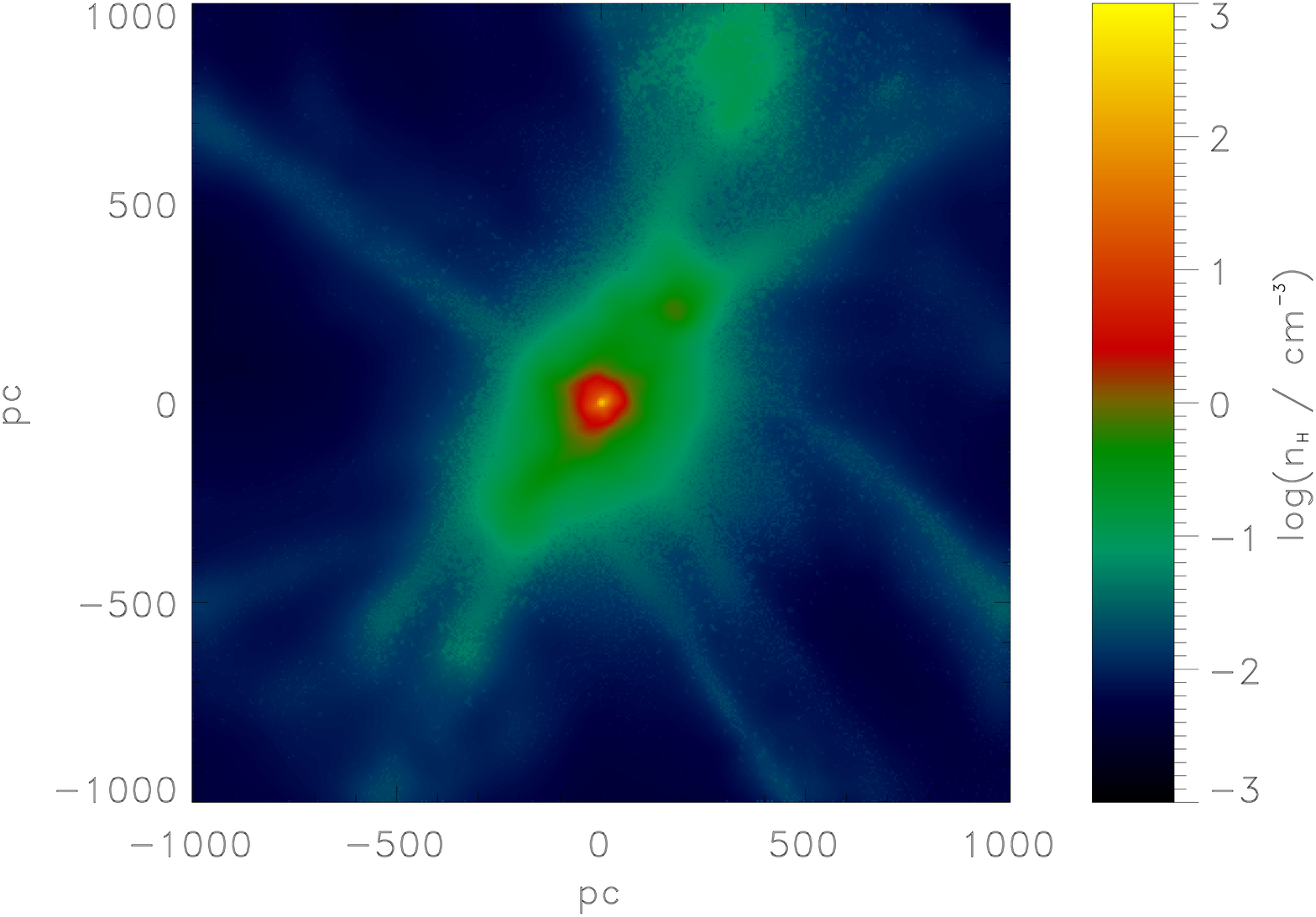}
\includegraphics[width=0.47\textwidth]{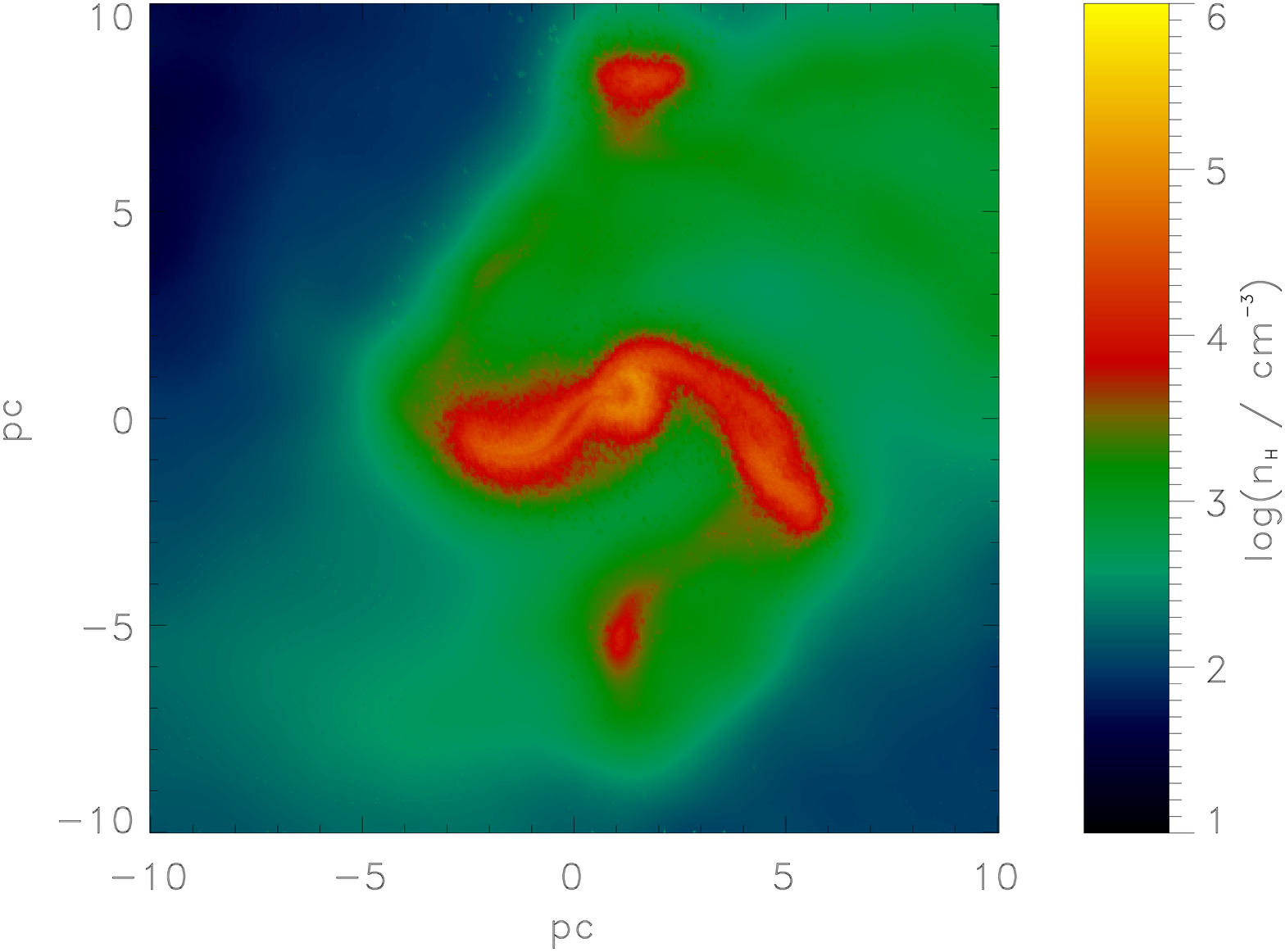}
\caption{Maps of the average number density of hydrogen nuclei along the line of sight for halo C with the \TreeCol approach at the moment of collapse ($z \sim 15.1$) at a scale of 2000\,pc (top) and 20\,pc (bottom). The background flux is $J_\mathrm{LW}=10^3 < J_\mathrm{crit}$, and we can clearly see the formation of clumps in the central region.}
\label{fig:cosmicweb}
\end{figure}

Altogether, we study the collapse of four different haloes (A, B, C, and D) with two different methods (\TreeCol and Jeans approximation) for determining the column density and several different strengths of the LW background per halo. All these runs are completely independent, and the column densities are calculated self-consistently during the simulation.
As an example, the structure of halo C and its central region is illustrated in Fig. \ref{fig:cosmicweb}.
We can see that the halo is embedded in the cosmic web and is fed by several gas streams. The central region shows a lot of substructure and gas clumps, which indicate ongoing fragmentation. In this case, the LW radiation is not strong enough to prevent efficient \h2 cooling and the gas can locally contract before the cloud globally collapses.

\subsection{Determination of $\Jcrit$}
\begin{figure}
\centering
\includegraphics[width=0.47\textwidth]{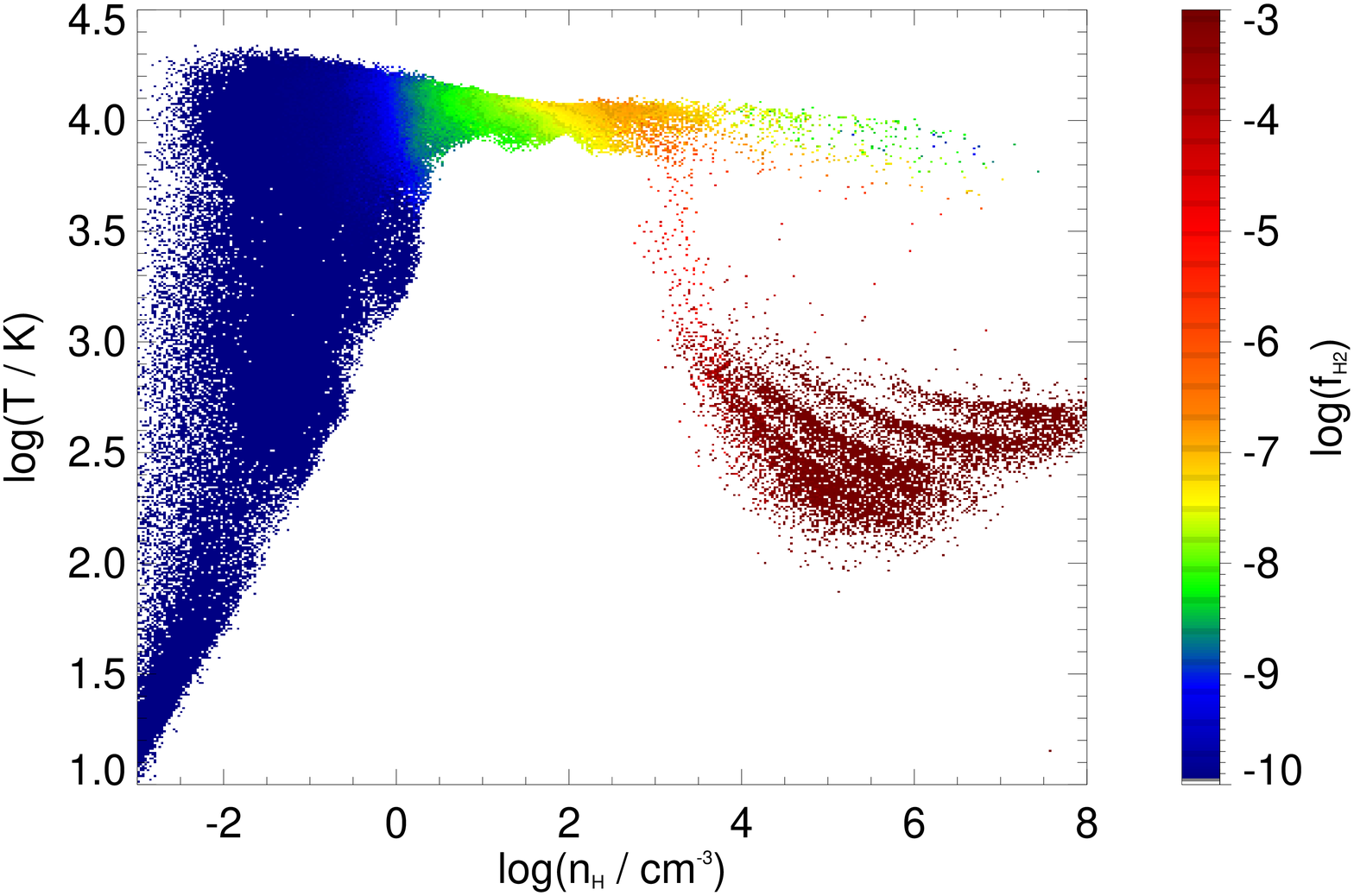}
\includegraphics[width=0.47\textwidth]{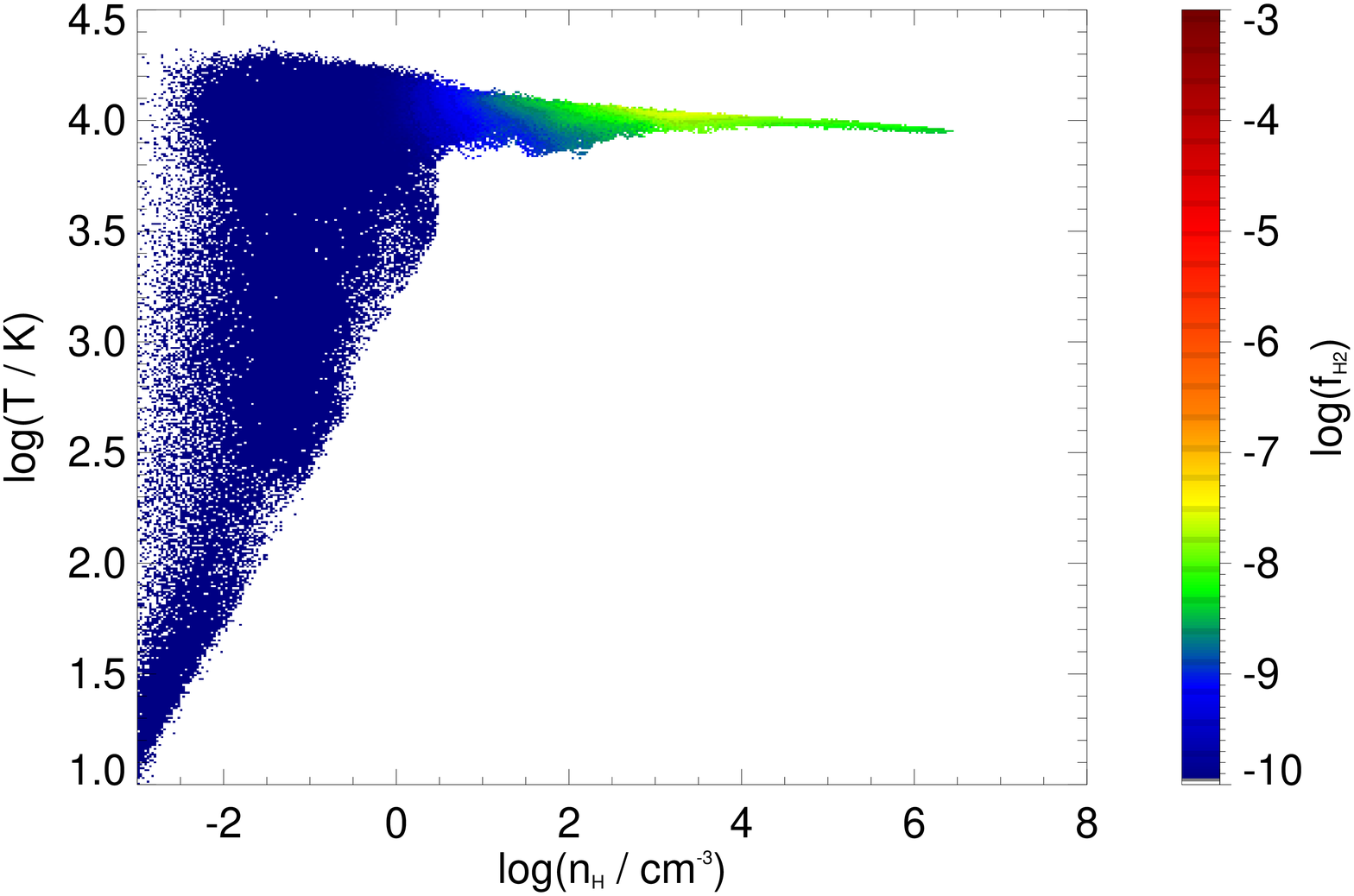}
\caption{Temperature as a function of number density of hydrogen nuclei colour-coded by the H$_2$ fraction for halo C using \TreeCol with $J_\mathrm{LW}=10^3$ (top) and $J_\mathrm{LW}=10^4$ (bottom). To better illustrate the relevant difference at higher \h2 fractions, we artificially set a lower limit of $f_{\mathrm{H}_2} = 10^{-10}$ in these plots. We can clearly see the different collapse behaviours depending on the strength of the LW background. With a high $J_\mathrm{LW}$, the gas remains hot around $10^4$\,K with $f_{\mathrm{H}_2} \lesssim 10^{-7}$. For a lower strength of the photodissociating background, the fraction of molecular hydrogen rises up to $f_{\mathrm{H}_2} \simeq 10^{-3}$ and the gas cools down to several hundred K. In the latter case, small clumps decouple from the global thermal evolution and we can see their footprints as stripe-like structures in the cold high-density gas.}

\label{fig:Cn_temp}
\end{figure}

The value of $\Jcrit$ sets the threshold between the two different collapse regimes. Above this value, the \h2 fraction remains low, the temperature stays around $10^4$\,K during the collapse, and only one central density peak forms. Below this value, the photo-dissociating background is not strong enough and \h2 line emission cools the collapsing gas to a few hundred K. This collapse typically results in several fragments \citep{clark11a,clark11b,greif11,stacy12,hirano14}. In order to discriminate between these two scenarios, we have to analyse the temperature evolution during the collapse. For two typical cases, the phase diagrams are given in Fig. \ref{fig:Cn_temp}.
During the virialisation of the halo, the gas shock heats to the virial temperature of around $10^4$\,K. The gas contracts further and remains at this temperature due to cooling by atomic hydrogen. With increasing density, also the column density of \h2 increases and the self-shielding against the external radiation becomes more efficient. As discussed above, H$_2$ reaches LTE at $n \simeq 10^{4}\,\mathrm{cm}^{-3}$, and if the collapse is still isothermal up to these densities, it will proceed isothermally. If the LW background is not strong enough, the \h2 fraction can increase, which in turn increases the column densities and hence leads to a more efficient self-shielding, which consequently increases the \h2 fraction even further. This runaway production of \h2 enables a clear distinction between the two different collapse regimes. An \h2 fraction of $\sim 10^{-3}$ is sufficient to cool the gas to temperatures of a few hundred K and hence to induce gas fragmentation. The individual clumps that form in this latter scenario can also be seen in the phase diagram as stripes in the cold, high-density regime, indicating that their thermal evolution is decoupled from one another.

\begin{figure}
\centering
\includegraphics[width=0.47\textwidth]{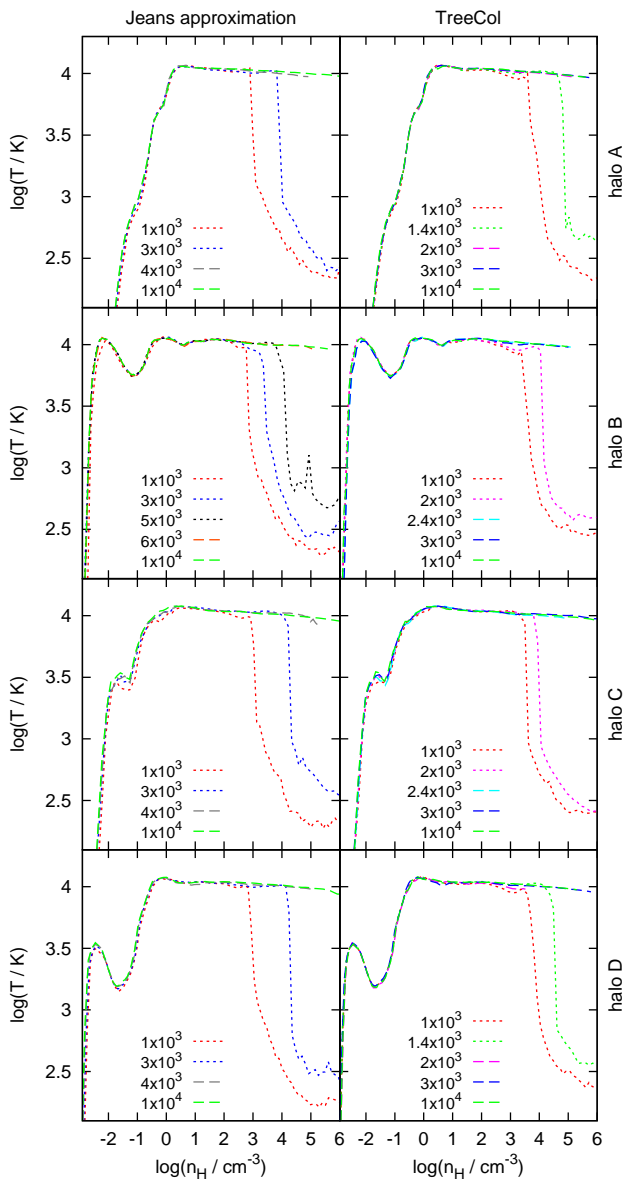}
\caption{Temperature as a function of the number density of hydrogen nuclei for the four haloes and the two different methods for determining the \h2 column density. The curves represent several realisations with different LW backgrounds, where the long-dashed lines represent the isothermal collapse and the short-dashed lines the collapse with efficient \h2 cooling. From these plots, we can read off the critical value for the isothermal collapse, which is systematically lower for the runs based on the \TreeCol method.}
\label{fig:n_temp}
\end{figure}

In order to find the values of $\Jcrit$, we have to compare the phase diagrams for different values of the LW background. To be able to do so, we bin and plot the data in logarithmic density space. An iterative algorithm merges and splits these density bins until each bin contains roughly the same number of Voronoi cells. This procedure ensures a statistically representative binning, but can have the side effect that the actual peak density in the simulation is not the same as the one in the binned data. The corresponding plots for the four haloes and the two different methods are shown in Fig. \ref{fig:n_temp}.
We test and display the background strengths of $J_\mathrm{LW}=10^3, 3\times 10^3, 10^4$ for all haloes and then successively bracket the actual value of $\Jcrit$.
We stop the simulations after the first snapshot with a peak density of $n \geq 10^6\ccm$ and compare the temperatures in the density regime above $n \geq 10^4\ccm$. If the temperature falls below 6000\,K, the collapse is regarded as non-isothermal. With this method, we find the lowest value $J_1$ for which the collapse is still isothermal and the highest value $J_2$ for which \h2 can efficiently cool the gas. Due to a limited number of possible realisations, we define the final critical value as the geometrical mean between these two values
\begin{equation}
\Jcrit = \sqrt{J_1 J_2}.
\end{equation}
Due to this finite number of tested $J_\mathrm{LW}$, the final values have an uncertainty of $\sim 10\%$. The resulting values, the virial masses of the haloes, and the collapse redshifts are compared in Table \ref{tab1}.
\begin{table}
 \centering
 \begin{tabular}{|c|c|c|c|c|}
  halo & $\Jcrit$ (Jeans) & $\Jcrit$({\sc treecol}) & $M_\mathrm{vir} / 10^7 \mathrm{M}_\odot$ & $z_\mathrm{coll}$\\ 
  \hline 
  A & 3500 & 1700 & 1.8 & 17.9\\ 
  \hline 
  B & 5500 & 2200 & 1.2 & 14.4\\ 
  \hline 
  C & 3500 & 2200 & 1.7 & 15.1\\ 
  \hline 
  D & 3500 & 1700 & 1.1 & 12.9\\ 
  \hline 
  \end{tabular}  
  \caption{Critical values $\Jcrit$ of the LW background for the four different haloes and the two different column density approaches. The $\Jcrit$ determined with the \TreeCol method is smaller by about a factor of two in all haloes. The halo-to-halo variance of this value is small and the collapse redshifts are distributed in a reasonable range. We also list the virial mass and the collapse redshift, which indicates the time when the halos first reach a density of $n \geq 10^6\ccm$.}
   \label{tab1}
\end{table}
First of all, we directly see that $\Jcrit$ is about a factor of two lower in the runs where we use the \TreeCol method to calculate the column densities. The reasons for this effect will be discussed in detail below. However, already the results based on the commonly used Jeans approximation are lower than found in previous studies. There are two main reasons for this. First, e.g. \citet{latif14a} and \citet{sugimura14} show that the self-shielding function by \citet{wolcott11} yields values of $\Jcrit$ that are up to an order of magnitude lower, compared to those derived with the function by \citet{draine96}. Secondly, the {\sc enzo} chemical model, which was used by many studies in this field \citep[e.g.][]{shang10,wolcott11,regan14b,latif14a}, tends to overestimate $\Jcrit$ by about a factor of two \citep{glover15a}. Hence, our results based on the Jeans approximation are in compliance with previous studies.

\subsection{Differences in the \h2 self-shielding}
In order to understand the differences induced by the new \TreeCol approach, we have to compare the column densities and corresponding self-shielding factors for the two methods. First of all, we summarise the most important features of this new method to understand these results. With \TreeCol, we create a spherical grid with 48 pixels around each Voronoi cell and project the column densities on to this grid. However, we do not calculate the total column densities, because gas can only contribute to the self-shielding if its relative velocity is smaller than $\sim 1.7$ times the thermal velocity (equation \ref{eq:hart15}). Based on this criterion we determine the spherical maps of effective column densities, which represents the spatial distribution of the self-shielding gas. The self-shielding factors are then calculated based on the column density for each of the pixels separately and then averaged over the 48 directions. This procedure is physically motivated, because the product $J_\mathrm{LW} f_\mathrm{sh}$ in equation (\ref{eq:dissrate}) represents an effective photodissociating flux and we simply average these fluxes over 48 different directions.

\begin{figure}
\centering
\includegraphics[width=0.47\textwidth]{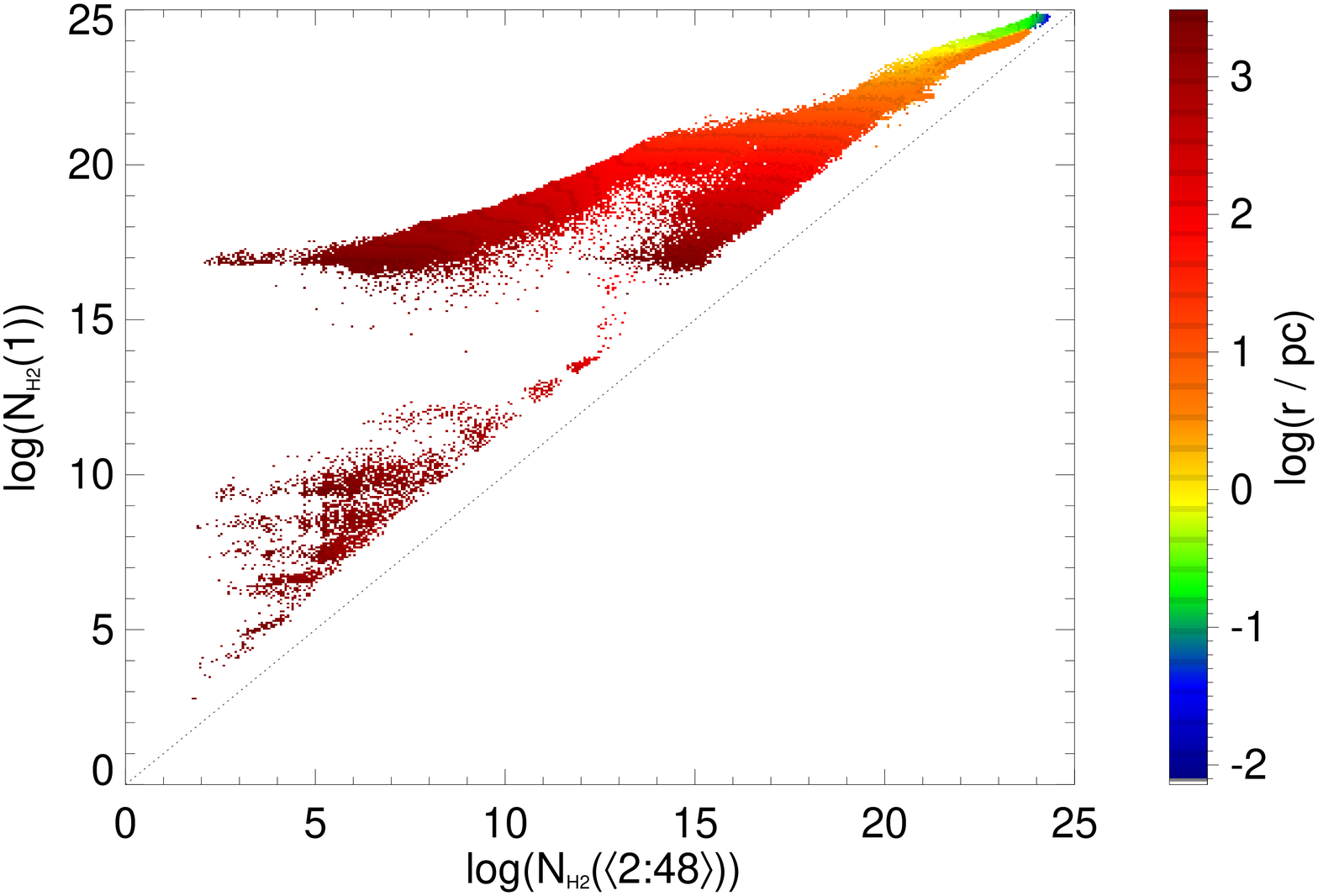}
\includegraphics[width=0.47\textwidth]{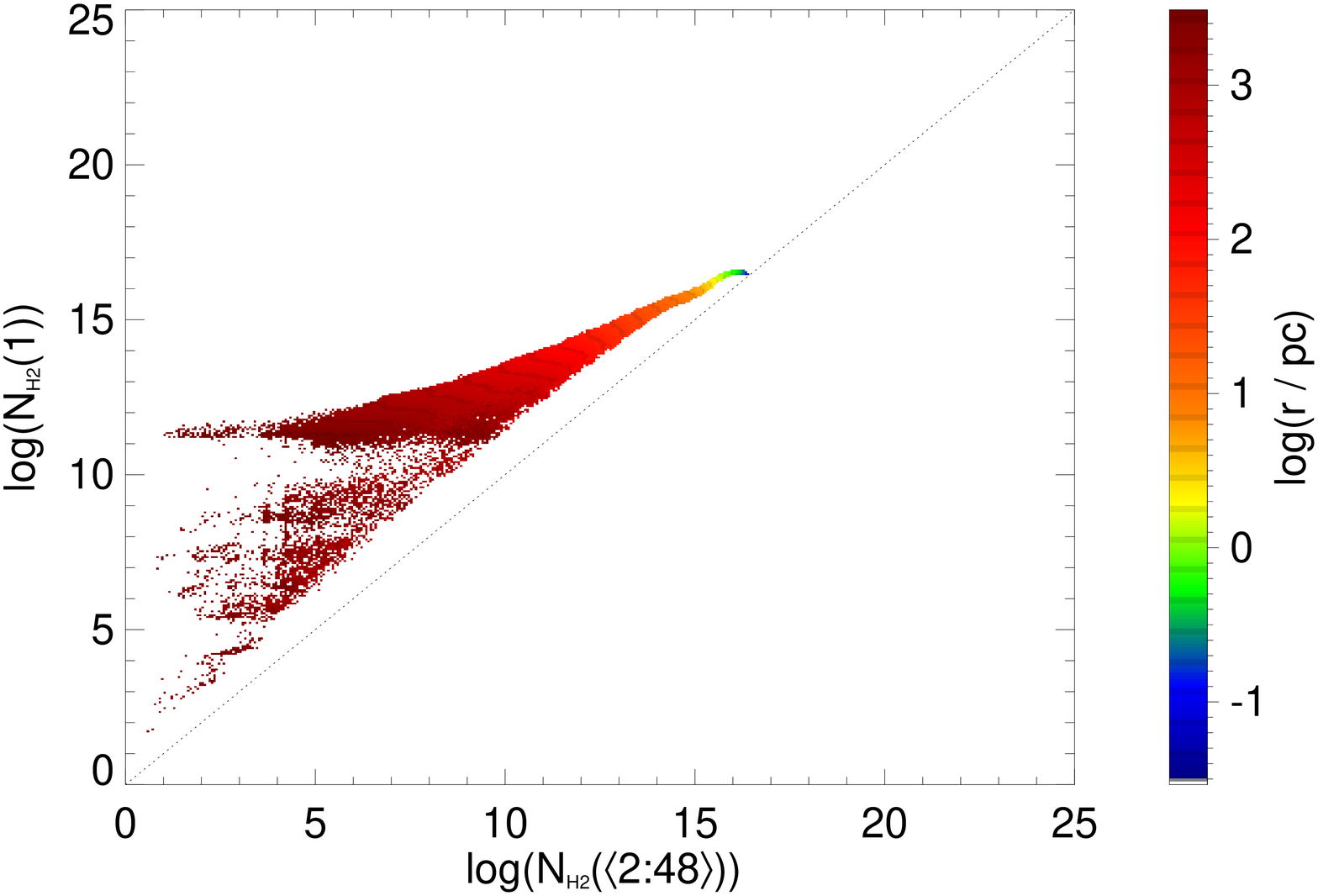}
\caption{Plots of the maximal column density for each Voronoi cell $N_{\mathrm{H}2}(1)$ as a function of the mean \h2 column density averaged over the remaining 47 pixels $N_{\mathrm{H}2}(\langle 2:48 \rangle)$, colour-coded by the radius. In the top panel, we show the results for halo C and a flux of $J_\mathrm{LW}=10^3<\Jcrit$ and in the lower panel for the same halo but with $J_\mathrm{LW}=10^4>\Jcrit$. This plot illustrates the huge directional dependence of the column densities, because for one Voronoi cell, the column densities in different directions vary by up to ten orders of magnitude. Interestingly, we can also see the more spherically symmetric collapse structure in the isothermal case (lower panel), because the plot converges towards the diagonal for small radii, indicating that the column densities are distributed isotropically close to the centre. Whereas in the upper panel, the angular distribution of column densities remains anisotropic even in the central region.}
\label{fig:colminmax}
\end{figure}

\begin{figure}
\centering
\includegraphics[width=0.47\textwidth]{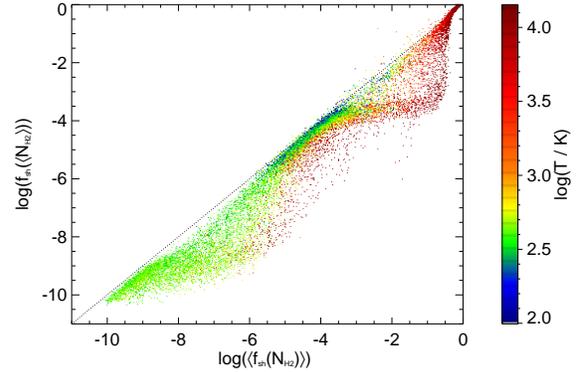}
\caption{Self-shielding factor for Halo C with the \TreeCol approximation and $J_\mathrm{LW}=10^3<\Jcrit$, colour-coded by the gas temperature. This plot illustrates the directional dependence of the self-shielding factor. In our simulation, we determine individually the self-shielding factors for the 48 different directions and average them afterwards (horizontal axis of this plot). On the vertical axis we see the self-shielding factor for exactly the same simulation output, but here, we first average the 48 different column densities for each fluid element and calculate the self-shielding factor based on this one averaged column density. The proper direction-dependent treatment of the H$_2$ self-shielding generally yields a less efficient shielding against LW radiation. This discrepancy is smaller for temperatures of a few hundred K.}
\label{fig:f_f}
\end{figure}

The importance of this point becomes clear, if we analyse the directional dependence of the column densities in Fig. \ref{fig:colminmax}.
We see that the column density distribution around one Voronoi cell is generally highly asymmetric and dominated by the contributions from one direction. In the case of a collapsing halo it is certainly the central high-density peak that yields the strongest contribution to the column density. From this direction, obviously, we should not expect any photodissociating radiation. A different way of presenting this important directional dependence of the self-shielding factor is given in Fig. \ref{fig:f_f}.
Averaging the column densities first and then calculating the shielding factor based on the one mean column density is highly biased by the contribution of one dominating direction. Consequently, it is important to properly average the effective photodissociating fluxes, as we do in our simulation. In contrast, a local approximation that only yields one column density might be affected by the central density peak and consequently underestimate the self-shielding factor.

\begin{figure}
\centering
\includegraphics[width=0.47\textwidth]{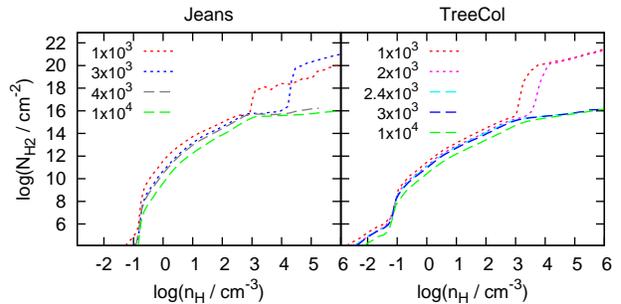}
\caption{Molecular hydrogen column density as a function of the density of hydrogen nuclei for halo C. For the \TreeCol runs, we plot the median column density over the 48 pixels in order not to be biased by one dominating direction. The values of the column density are calculated self-consistently during the simulation. Hence, they are not directly comparable for the two different approaches, because the structure of the cloud might be different. However, we can already see important similarities such as a threshold column density of about $10^{16}\,\mathrm{cm}^{-2}$ above which strong self-shielding enables the formation of sufficient \h2 to cool the cloud efficiently.}
\label{fig:n_colC}
\end{figure}

\begin{figure}
\centering
\includegraphics[width=0.47\textwidth]{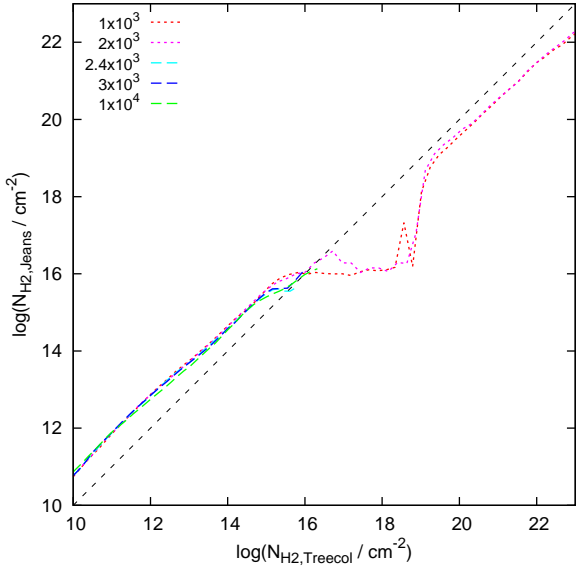}
\caption{Direct comparison of the \h2 column densities for the two different approaches. The data presented here is based on an output of halo C simulated with \TreeCol and the corresponding column densities based on the Jeans approximation are calculated based on the output file. Again, we use the median over the 48 pixels for the column densities based on \TreeCol. For column densities below $10^{16}\,\mathrm{cm}^{-2}$, the Jeans approximation yields column densities that are higher by about one order of magnitude.}
\label{fig:col_colC}
\end{figure}

Now, we want to compare directly the results for the column density and the shielding factor based on the two implementations. In Fig. \ref{fig:n_colC}, we display the column densities of the simulations with the Jeans approximation and the \TreeCol approach.
These column densities are calculated self-consistently during the run and are therefore not directly comparable. However, we already see that the column densities in the isothermal case remain under $10^{16}\,\mathrm{cm}^{-2}$ and that, for smaller values of $J_\mathrm{LW}$, the column density increases already for lower densities. To be able to directly compare the effect of the two different approaches, we use one snapshot of the simulations based on \TreeCol and determine the corresponding Jeans column densities by post-processing the data. The comparison of these column densities is given in Fig. \ref{fig:col_colC}.
In the isothermal case, the column densities remain below $10^{16}\,\mathrm{cm}^{-2}$ and in this regime, the Jeans approximation overestimates the column densities by approximately one order of magnitude. Consequently, the shielding in this regime is more efficient with the Jeans approximation, and we need a higher LW background to obtain an isothermal collapse. For column densities above $10^{16}\,\mathrm{cm}^{-2}$, the Jeans approximation underestimates the values. We find that these regions, where the Jeans approximation underpredicts the column densities are the lower density regions between the clumps (compare e.g. with the lower panel in Fig. \ref{fig:cosmicweb}). In this low-density environment, the Jeans approximation sees only the local gas conditions and predicts a rather small column density, whereas \TreeCol is able to capture the nearby high-density clumps with high \h2 fractions. Consequently, \TreeCol (correctly) yields higher column densities in this regime. However, this underestimation does not affect the value of $\Jcrit$, because we only get \h2 column densities significantly above $10^{16}\,\mathrm{cm}^{-2}$ in haloes where $J_\mathrm{LW} < \Jcrit$ and where the gas can undergo runaway cooling. Phrased differently, this underestimation by the Jeans approximation at higher column densities is a consequence of the runaway \h2 cooling and the subsequent fragmentation and not its trigger.

There are two obvious reasons why \TreeCol yields lower effective column densities. First of all, it takes into account the three-dimensional distribution of the matter and the declining density radially outwards, whereas the Jeans approximation assumes a constant density within one Jeans length and hence generally overestimates the gas number density. Secondly, only fluid elements within a certain velocity range contribute to the effective column density in \TreeCol, which again reduces the value of the column density. \citet{wolcott11} have already pointed out that the Jeans approximation generally overestimates the column densities in a static, isothermal slab of gas. Our treatment of the relative velocities further increases this effect.

\begin{figure}
\centering
\includegraphics[width=0.47\textwidth]{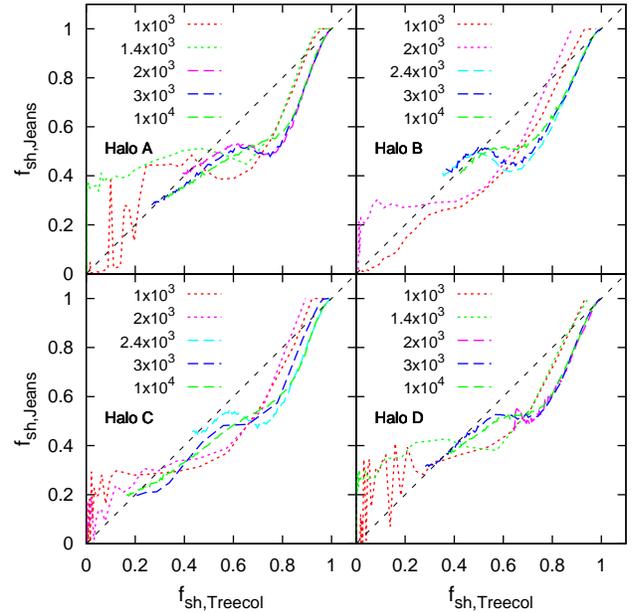}
\caption{Direct comparison of the self-shielding factors for the two different approaches. The data presented here is based on haloes simulated with \TreeCol and the corresponding Jeans column densities are calculated based on the output files. In the isothermal regime (long-dashed lines), the \TreeCol approximation yields higher self-shielding factors and therefore provides a less efficient shielding against the external photodissociating radiation field.}
\label{fig:f_f2}
\end{figure}

We can see the same trend for the self-shielding factors in Fig. \ref{fig:f_f2}.
In the case of an isothermal collapse, the \TreeCol methods yields always higher values for the shielding factor and therefore a less efficient shielding against the LW background. Consequently, a smaller value of $\Jcrit$ provides already enough photodissociating radiation to keep the collapse isothermal.

\subsection{Impossibility of a simple correction factor}
Although \TreeCol is computationally efficient, it would be good if there is a simple correction formula to the Jeans approximation that is able to reproduce the self-shielding results obtained from \TreeCol at least to some degree. The aim would be to find an updated version of equation (\ref{eq:db96}) that uses the column densities $N_{\mathrm{H}_2} = n_{\mathrm{H}_2} L_\mathrm{J}$ based on equation (\ref{eq:Jeans}). As a start, we can impose the velocity criterion from equation (\ref{eq:hart15})  on the matter included in the calculation of the effective \h2 column density. This changes the dependence of the self-shielding factor on the thermal velocity, which is expressed by the exponent $\alpha$ in equation (\ref{eq:db96}). Hence, we try to fit the results obtained with \TreeCol with the free parameter $\alpha _*$ but with the column densities determined with the Jeans approximation as an input:
\begin{equation}
f_\mathrm{sh}(N_{\mathrm{H}_2,\mathrm{TreeCol}})_{\alpha=1.1} = f_\mathrm{sh}(N_{\mathrm{H}_2,\mathrm{Jeans}})_{\alpha _*}
\end{equation}
However, this exponent varies with a huge scatter between the different haloes and with the various strengths of the background radiation. For example, for halo C we find $\alpha _* = 0.6$ for $J_\mathrm{LW} = 2400$ and $\alpha _* = 1.2$ for $J_\mathrm{LW} = 4000$, although both collapse isothermally. Most other values are distributed in the range $0.7 < \alpha _* < 1.1$, which illustrates that it is not possible to find a simple correction factor to reproduce the results by \TreeCol. This is mainly based on the fact that a proper determination of the self-shielding factor has to take into account the three-dimensional structure of the density and velocity field, as we have seen before. Since the Jeans approximation is only based on local quantities, it cannot capture this structure and consequently fails at reproducing the self-shielding factors.

\subsection{Effect of damping wings}
\label{sec:wings}
So far, we assumed that only matter in a certain velocity range contributes to the self-shielding of H$_2$, because for larger relative velocities the spectral lines are Doppler-shifted too far from the core. As already mentioned in section \ref{sec:chem}, this picture changes for higher column densities, where the contribution of the damping wings becomes important and different lines can self-shield each other \citep{black77, draine96}. This effective broadening of spectral lines makes relative velocities less important and the total H$_2$ column density then contributes to self-shielding. The original self-shielding formula by \citet{draine96}, equation \ref{eq:db96} in section \ref{sec:chem}, already included these two contributions, where the first term corresponds to shielding from the cores of individual lines, and the second term represents the effect of overlapping damping wings, which are not affected by relative velocities. \citet{gnedin14} propose a correction to the self-shielding factor of the form
\begin{equation}
f_\mathrm{sh} = \frac{0.965}{(1+x_1/b_5)^\alpha} + \frac{0.035}{(1+x_2)^{0.5}} \exp \left[ - \frac{(1+x_2)^{0.5}}{1180} \right],
\label{eq:gnedin14}
\end{equation}
where $x_1=N_{\mathrm{H}_2,\mathrm{treecol}} / 5 \times 10^{14} \mathrm{cm}^{-2}$ is the effective H$_2$ column density and $x_2=N_{\mathrm{H}_2,\mathrm{total}} / 5 \times 10^{14} \mathrm{cm}^{-2}$ is the total H$_2$ column density. \citet{gnedin14} use the exponent $\alpha=2$ based on the original work by \citet{draine96}, whereas we apply the more recent value $\alpha=1.1$ by \citet{wolcott11}.

\begin{figure}
\centering
\includegraphics[width=0.47\textwidth]{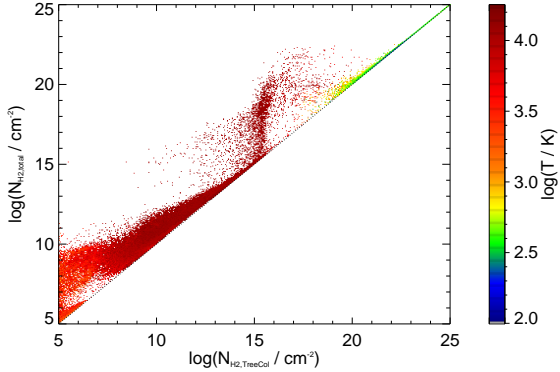}
\caption{Total \h2 column density as a function of the effective \h2 column density colour-coded by the gas temperature for halo C with $J_\mathrm{LW} = 1000 < J_\mathrm{crit}$. For the effective column density, we impose the velocity criterion of equation \ref{eq:hart15} and only matter fulfilling this criterion is included. The excess at $N_{\mathrm{H}_2,\mathrm{treecol}} \simeq 10^{16} \mathrm{cm}^{-2}$ is created by the hot gas, which falls on to the cold and dense gas clumps and has therefore a high relative velocity. Hence, the velocity criterion excludes a lot of matter, and the total column density is significantly higher in this regime.}
\label{fig:coltot}
\end{figure}

We investigate the impact of the damping wings by determining the total H$_2$ column densities. To do so, we post process the snapshots of halos C with \TreeCol and compare them to the effective column densities (Fig. \ref{fig:coltot}).
The total column density is higher by up to several orders of magnitude. Especially around $N_{\mathrm{H}_2,\mathrm{treecol}} \simeq 10^{16} \mathrm{cm}^{-2}$, we see that the total column density is much larger than the effective one. This excess is created by the hot gas that falls towards the centre and has therefore a high relative velocity with respect to these central cold and dense gas clumps. Consequently, the velocity criterion excludes a large contribution, and the total column density is significantly higher than the effective one in this regime. In any case, this comparison strengthens the importance and illustrates the influence of the velocity criterion on the determination of the column densities for self-shielding.

We now use these two column densities and determine the self-shielding factors with equation (\ref{eq:gnedin14}) and compare them to the previously used values based on equation (\ref{eq:db96}). In the case of $J_\mathrm{LW} = 10^4 > J_\mathrm{crit}$, the effective and the total column densities remain below $\sim 10^{16} \mathrm{cm}^{-2}$, and the contribution of damping wings is negligibly small: the mean deviation between the self-shielding factors is $0.24\%$ with a maximal discrepancy of $0.60\%$. This is due to the fact that the main contribution to the self-shielding in this regime comes from the cores of individual lines and hence, from the first term in equation (\ref{eq:gnedin14}). For the case with $J_\mathrm{LW} = 10^3 < J_\mathrm{crit}$, we have a higher \h2 abundance and consequently higher \h2 column densities. Here, the self-shielding factors differ on average by $2.2\%$ with a maximal deviation of $5.3\%$. Hence, we conclude that the contribution of the damping wings to the self-shielding is negligibly small in our scenario, compared to other effects and approximations. In particular, it seems not to affect the determination of $J_\mathrm{crit}$, because this effect only becomes important, once the runaway \h2 production has already set in.

\subsection{Mass infall rate}
\begin{figure}
\centering
\includegraphics[width=0.47\textwidth]{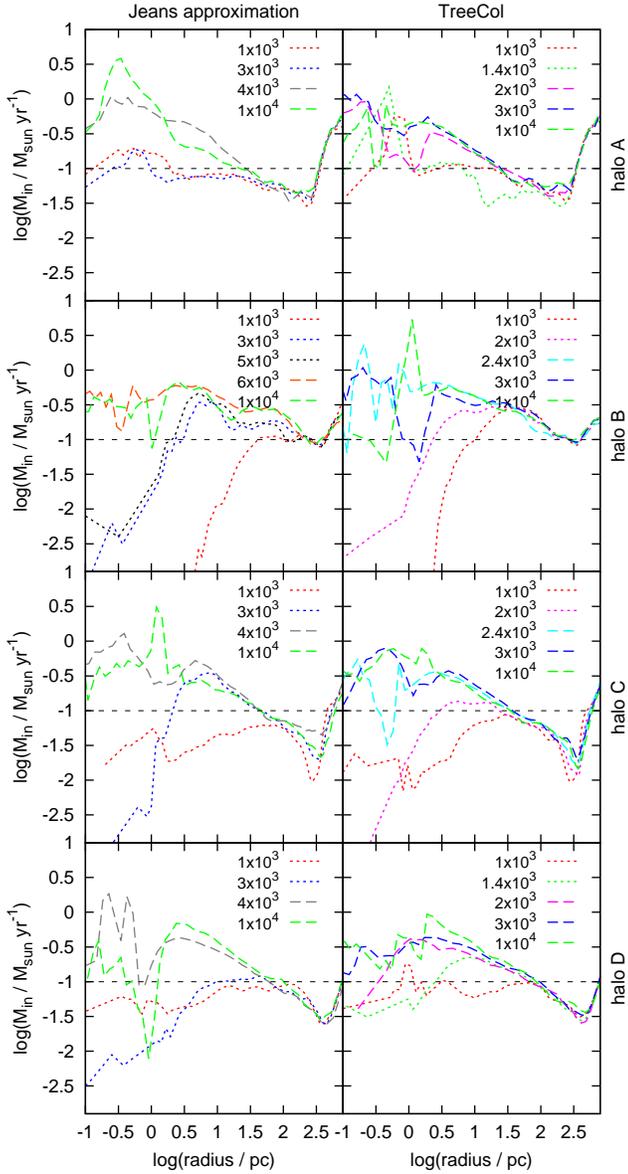}
\caption{Radial profiles of the mass infall rates averaged over the last $\sim 10^5$\,yr of the collapse. Simulations with $J_\mathrm{LW} < \Jcrit$ are shown with short-dashed lines and those with $J_\mathrm{LW} > \Jcrit$ are shown with long-dashed lines. The black line at $M_\mathrm{in} = 0.1\,\mathrm{M}_\odot \mathrm{yr}^{-1}$ represents the theoretical threshold above which the formation of a SMBH seed is possible. Generally, the mass infall rates are higher in the case of an isothermal collapse and especially in the central regions, the mass infall rates fall significantly below $0.1\,\mathrm{M}_\odot \mathrm{yr}^{-1}$ for $J_\mathrm{LW} < \Jcrit$.}
\label{fig:rMdot}
\end{figure}

The main quantity that determines if a SMS and hence a seed of a SMBH forms is the mass infall rate $M_\mathrm{in}$, which has to be above $M_\mathrm{in} \gtrsim 0.1 \mathrm{M}_\odot\,\mathrm{yr}^{-1}$ \citep{begelman10,hosokawa13,schleicher13,latif13a,ferrara14}.
An isothermally collapsing cloud at $T \simeq 10^4$\,K is a sufficient criterion to provide the necessary mass infall rates, but is this criterion also necessary? To study this question, and to see if the \h2 self-shielding implementation induces any difference, we analyse the mass infall rate for the different haloes and methods. Assuming a spherically symmetric cloud, we determine the mass infall rate by
\begin{equation}
M_\mathrm{in} = 4 \pi r^2 |v_r| \rho ,
\end{equation}
where $r$ is the distance from the densest point, $v_r$ is the radial velocity, and $\rho$ is the gas density. We average the radial velocity and the gas density over spherical shells. However, we note that the mass infall rate is generally not spherically symmetric and neither is it constant in time. We therefore take the mean over the last $\sim 10^5$\,yr of evolution to obtain a reasonably smooth and well-defined value (Fig. \ref{fig:rMdot}).
The mass infall rates in the isothermal collapse scenarios are in the range $0.1\,\mathrm{M}_\odot \mathrm{yr}^{-1} \lesssim M_\mathrm{in} \lesssim 1\,\mathrm{M}_\odot \mathrm{yr}^{-1}$ and tend to be higher than those in the cases with $J_\mathrm{LW} < \Jcrit$. Especially in the central $\sim 10$\,pc, only the isothermal clouds are able to provide mass infall rates above $0.1\,\mathrm{M}_\odot \mathrm{yr}^{-1}$, which are necessary for the formation of SMBH seeds. As we have seen before, the self-shielding method influences the value of $\Jcrit$, but if we now compare the regimes below and above $\Jcrit$ separately for both approaches, we do not see any clear systematic difference that is induced by the choice of the self-shielding method.

\begin{figure}
\centering
\includegraphics[width=0.47\textwidth]{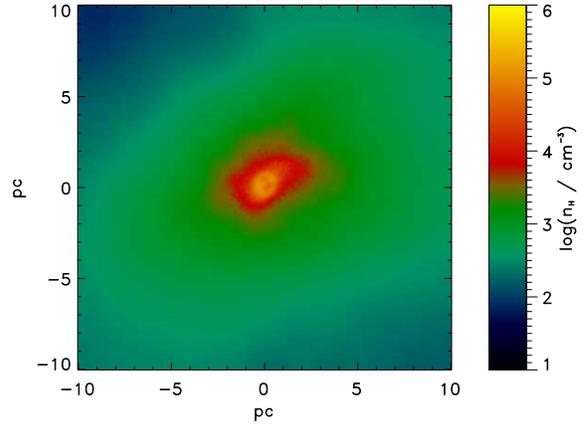}
\caption{Map of the average number density of hydrogen nuclei along the line of sight for halo C with the \TreeCol approach at the moment of collapse ($z \sim 15.1$). The background flux is $J_\mathrm{LW}=3 \times 10^3 > J_\mathrm{crit}$ and in contrast to Fig. \ref{fig:cosmicweb}, we can see no indication for gas fragmentation but a rather smooth, approximately spherically symmetric accretion towards the centre.}
\label{fig:isothermalmap}
\end{figure}

Assuming that approximately one Jeans mass per free fall time falls towards the central region, the mass accretion rate is given by
\begin{equation}
\dot{M} = \frac{c_s ^3}{G} \propto T ^{3/2},
\end{equation}
where $c_s$ is the sound speed and $G$ the gravitational constant. Consequently, the accretion rate is higher for hotter gas, such as in the isothermally collapsing cloud that retains temperatures of $\sim 10^4$\,K during the collapse. Moreover, the Jeans mass drops with the temperature and the gas is more susceptible to fragmentation in the case of efficient \h2 cooling \citep[see e.g.][]{clark11a,clark11b}. Once cooling becomes efficient and the gas temperature falls, the cooling time becomes shorter than the local freefall time and the gas can locally contract, before the cloud globally collapses. The resulting clumpy structure can be seen in Fig. \ref{fig:cosmicweb}, whereas the spatial structure of the isothermal collapse is illustrated in Fig. \ref{fig:isothermalmap}.
This smooth, almost spherically symmetric structure without signs of fragmentation enables higher gas infall rates, which leads to the formation of a SMS which then collapses to a SMBH seed. However, one should keep in mind that the different mass infall rates for scenarios below and above $\Jcrit$ are just a trend and also the threshold of $0.1\,\mathrm{M}_\odot \mathrm{yr}^{-1}$ is only a rough estimator with other proposed values between $0.01$ and $1\,\mathrm{M}_\odot \mathrm{yr}^{-1}$ \citep{begelman10,hosokawa12,schleicher13,ferrara14}. We find $M_\mathrm{in} < 0.1\,\mathrm{M}_\odot \mathrm{yr}^{-1}$ also for the isothermal collapse and vice versa. A value of $0.6\,\mathrm{M}_\odot \mathrm{yr}^{-1}$ seems to discriminate between the two collapse regimes: all isothermally collapsing clouds yield accretion rates above this value, but only one out of eight clouds with $J_\mathrm{LW} < \Jcrit$ reaches this accretion rate.

Recently, \citet{latif15b} study the infall rates in atomic cooling haloes in greater detail. As in our study, they find that it is not always necessary to completely suppress \h2 formation to obtain sufficiently large infall rates to form a SMS. Moreover, they detect a rotationally supported structure in the central parsec, but this rotational support does not halt the collapse and still enables infall rates of $\sim 0.1\,\mathrm{M}_\odot \mathrm{yr}^{-1}$.
For a more detailed discussion of this topic, we refer the interested reader to \citet{latif15b}.

\section{Caveats}
\label{sec:caveats}
The attempt to find one universal value of $\Jcrit$ is rather artificial, because the relevant physical processes are too complex to be summarised in one simple number that decides whether we form a SMBH seed or not. Recently, \citet{agarwal15b} study the value of $\Jcrit$ using one-zone models with a more realistic spectral energy distribution for the external radiation and show that $\Jcrit$ is not one fixed value, but rather a range spanning more than three orders of magnitude. Moreover, we can also have sufficiently high mass infall rates to form a SMS even for $J_\mathrm{LW}<\Jcrit$. Although the concept of one universal threshold $\Jcrit$ is questionable in the formation scenario of SMBH seeds, it is a convenient quantification to study the influence of different physical processes on the direct collapse scenario.

\subsection{Stellar spectrum}
One should keep in mind that we use the T5 spectrum to study the effect of \h2 self-shielding (section \ref{sec:chem}). This is an accurate approximation for stars with a characteristic mass of $\sim 100\,\mathrm{M}_\odot$, but subsequent stellar populations are believed to have softer stellar spectra and might therefore yield a lower value of $\Jcrit$ \citep{sugimura14,agarwal15a,agarwal15b}. However, \citet{latif15a} show that the value of $\Jcrit$ only weakly depends on the adopted radiation spectra in the range $2 \times 10^4\,\mathrm{K} \leq T_\mathrm{rad} \leq 10^5\,\mathrm{K}$.

\subsection{Resolution}
\begin{figure}
\centering
\includegraphics[width=0.47\textwidth]{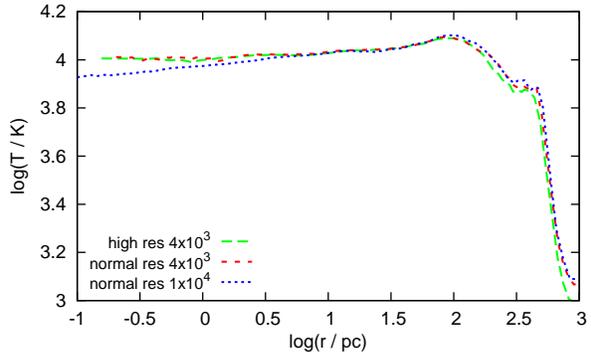}
\caption{Temperature profiles for halo C using the Jeans approximation, where the green long-dashed curve represents an additional run with a two times higher spatial resolution. This should be compared to the red short-dashed curve, which represents a run with the normal resolution and the same background flux. The blue dotted curve shows the profile for the same halo with normal resolution but a higher LW flux. All curves show an isothermal collapse, but the differences induced by a different background radiation are larger than those induced by a higher resolution.}
\label{fig:n_temp_res}
\end{figure}

To test the resolution, we also perform one simulation with a better mass resolution. In our standard approach, we resolve the local Jeans mass by 66 Voronoi cells, which correspond to a spatial resolution of $\lesssim 0.1$pc. For the high-resolution run, we increase the mass resolution by a factor of eight, which doubles the spatial resolution. The comparison can be seen in Fig. \ref{fig:n_temp_res}.
The run with a higher spatial resolution yields the same temperature profile as the run with the normal resolution. Hence, we can conclude that our normal resolution is high enough to properly resolve the collapse, and our results are not sensitive to the numerical resolution.

\subsection{Photochemistry}
The \citet{wolcott11} self-shielding function that we use here is intended for use when the H$_{2}$ is rotationally hot and not only the lowest rotational levels are populated. This is a reasonable approximation for densities $n \simeq 10^{3}$--$10^{4} \: {\rm cm^{-3}}$, but at lower densities we would expect most of the H$_{2}$ molecules to be in the $J = 0$ or $J = 1$ levels, and in this regime the \citet{wolcott11} self-shielding function will underestimate the effectiveness of H$_{2}$ self-shielding. The effect of this on $J_{\rm crit}$ is unclear, and we intend to investigate this further in future work. In addition, our current treatment of the shielding of H$_{2}$ does not account for absorption by the Lyman series lines of atomic hydrogen \citep{haiman97}. \citet{wolcott11b} show that shielding of \h2 by atomic hydrogen becomes important for column densities of $N_\mathrm{H} > 10^{23} \mathrm{cm}^{-2}$. Including this effect may yield larger values for $J_{\rm crit}$ and may hence exacerbate the difference between the Jeans approximation results and the results derived using {\sc treecol}.

\section{Conclusions}
\label{sec:summary}
We have implemented a new method for the determination of \h2 column densities in the 3D moving mesh code {\sc arepo} and used it to study the effect of an improved treatment of \h2 self-shielding on the `direct collapse' scenario. In a comparison to the previously used Jeans approximation, we find that the effective column densities are generally smaller with our new method and the necessary LW background flux to suppress efficient \h2 cooling is lower by a factor of about 2. More precisely, we find $\Jcrit \simeq 2000$ with our new approach compared to $\Jcrit \simeq 4000$ with the Jeans approximation. The main reason for this difference is the large directional dependence of the self-shielding factor that cannot be captured with one-dimensional methods. Because the detailed morphological and kinematic structure of the cloud matters a lot for the determination of the effective column density, it is also not possible to find a simple correction factor that might reproduce the results based on \TreeCol.

Following \cite{inayoshi14c}, the density of possible direct collapse black hole formation sites scales with $n_\mathrm{DCBH} \propto \Jcrit ^{-5}$ for $J_\mathrm{LW} > 10^3$. Consequently, the factor of two, by which $\Jcrit$ is lower with our new self-shielding approach, leads to a number density of direct collapse black holes in the early Universe that is about $32$ times higher than previously expected. Although the number of expected direct collapse black holes is significantly higher with our new method, the value of $\Jcrit \simeq 2000$ is still too high to explain the number density of SMBH at redshift $z \simeq 6$ of $n_\mathrm{SMBH} = 10^{-9}\,\mathrm{Mpc}^{-3}$ (comoving units) only by the isothermal direct collapse scenario. Even under optimistic assumptions, $\Jcrit$ has to be smaller than $1000$ to explain the observed number density \citep[see e.g.][]{dijkstra14,inayoshi14c}.

\subsection*{Acknowledgements}
The authors would like to thank M\'elanie Habouzit, Yohan Dubois, Joseph Silk, Gary Mamon, St\'ephane Colombi, Rebekka Bieri, Andrea Negri, Mitch Begelman, Bhaskar Agarwal, Britton Smith, Mei Sasaki, and Paul Clark for valuable discussions and helpful contributions. We also thank the referee for insightful comments. We are very grateful for the free availability of the code {\sc music}.
The authors acknowledge funding from the European Research Council under the European Community’s Seventh Framework Programme (FP7/2007-2013) via the ERC Grant `BLACK' under the project number 614199 (TH, MAL, MV) and via the ERC Advanced Grant `STARLIGHT: Formation of the First Stars' under the project number 339177 (RSK, SCOG).
SCOG and RSK acknowledges support from the DFG via SFB 881, `The Milky Way System' (sub-projects B1, B2 and B8) as well as via SPP 1573 `Physics of the Interstellar Medium'. The simulations described in this paper were performed at the J\"ulich supercomputing centre.

\bibliographystyle{mn2e}

\bibliography{H2SelfShielding}

\bsp

\label{lastpage}

\end{document}